\documentclass{article} 
\usepackage{nips15submit_e,times}
\usepackage{hyperref}
\usepackage{url}
\usepackage{amssymb}
\usepackage{amsmath}
\usepackage{graphicx}
\usepackage[T1]{fontenc}
\usepackage[font=small,labelfont=bf,tableposition=top]{caption}

\DeclareCaptionLabelFormat{andtable}{#1~#2  \&  \tablename~\thetable}

\title{Measuring Responsiveness in the Online Public Sphere for the 2016 U.S. Election}

\author{
Pau Perng-Hwa Kung\\
Media Lab\\
Massachusetts Institute of Technology\\
Cambridge, MA 01239 \\
\texttt{pernghwa@media.mit.edu} \\
\And
Deb Roy \\
Media Lab\\
Massachusetts Institute of Technology\\
Cambridge, MA 01239 \\
\texttt{dkroy@media.mit.edu} \\
}

\nipsfinalcopy 

\begin{document}

\maketitle

\begin{abstract}
The election narrative is formed under the competitions of ideas among critical players involving politicians, news media, public influentials, and the general public. Untangling the complex process of narrative formation, however, is no easy task due to implicit influences among the key players. In this paper, we propose the problem of measuring "responsiveness" that quantifies a player's influence on another given a specific election topic over time. In particular, we make use of multivariate Hawkes Process to infer the influence network between pairs of election players. We demonstrate an early version system of analytic pipeline of online public sphere discussions from data ingestion, influence inference, to visualization. The paper concludes by showcasing some preliminary results based on Twitter and news media election-related contents from July to October 2015 and discussing plans for future research.
\end{abstract}

\section{Introduction}

The 2016 U.S. Election is quite different from the previous elections: we have a more mature online public sphere than ever before. Recalling the German philosopher Habermas, the public sphere is an area where individuals gather to freely discuss on public issues[6]. With the prevalence of online social media, people's opinions on public issues have become more transparent[4], and we believe it is now possible to track the formation of public narrative on a given election topic.

However, the dynamics that govern the election narratives are complex: in an election, there are politicians, news media, influential public, and the general public making up the campaign voices. Each voice wants to get the podium to express its own views and interests. In other words, the election process is essentially a competition of ideas. At any given time, each voice may have varying influence on another regarding a certain election topic. In order to better understand the political narrative, it is important to be able to determine who and what that lead to a particular voice's discussion on a certain topic. During a specific time window, is it possible to identify key opinion influencers and triggers that affect a voice to respond differently on an election topic? Developing a computational understanding can provide the general public a clearer picture on which politician or news organization has aligned topic interest. Knowing how a candidate is responsive to some candidates under certain contexts but not others can be quite informative. 

In this paper, we develop a computational measure of narrative influence in the online public sphere, called "responsiveness". Essentially, we model each voice's reactions on a particular election topic as a point process, where a voice can be an individual, organization, or a group of individuals with aligned interest. We believe the topical influence acts like ripple effects among the voices. Based on this assumption and jointly considering voices as multiple point processes, we model the cross-voice influence using an excitatory point process called multivariate Hawkes Process. We describe a solution framework entailing our data collection process, algorithm inference, and outcome visualization. A quick snapshot of the initial version of the system is shown, focusing on election-related tweets and news articles collected over the period of July thru October in 2015. We illustrate some initial application areas using the inferred influence between voices over time including interest alignment analysis, situational analysis, and topic shift analysis. We conclude by discussing potential improvements of the current algorithm and application cases.

\section{Problem and Solution Framework}
\subsection{Preliminaries and Problem Formulation}
Suppose given a specific content filtering criteria $C$, where the filtering criteria may be discussions regarding certain election topics (e.g. abortion) or candidates. And let the number of election voices we wish to analyze be $K$. We observe $K$ series of events, where each series $i$ represents event process belonging to voice $i$. Events can be represented as a finite sequence of marked tuples $\{s_n,\ c_n\}_{n=1}^N$, where $c_n \in \{1,...,K\}$ denotes the process in which the $n$-th event occurred and $s_n \in [0,T]$ denotes the timestamp of the event. 

We define the measure Responsiveness $R: N$ x $N$ x $H \rightarrow$ $\mathbb{R}$ to be the following: responsiveness $R(s_i,t_i,h | C)$ is a function that measures influence from process $s_i$ to process $t_i$ given data filtering criteria $C$, where $s_i,t_i \in S,K=\{1,...,K\}$ denote the indices of the influencing process and the influenced process, respectively. $h \subseteq H = [1,T]$ is the time interval in which we want to measure the responsiveness. Therefore, the problem we face in this paper is that given any sequence of comments by a set of voices regarding any election topic, can we infer Responsiveness for pairs of voices over a certain time period?

In fact, this measure can be further augmented to measure nuanced signals associated with the process-to-process influence. For example, we can add an augmenting topic variable $\Gamma = \{1,...,Z\}$ denoting a set of election topics (e.g. $\{$[abortion], [immigration],..., [gun control]$\}$) so the Responsiveness can now be $R(s_i,t_i,h,\gamma | C)$, $\gamma \in \Gamma$. 

An example would be to let $C$ = discussion on immigration; $\gamma$ = [Mexican border]; $s_i$ = Donald Trump; $t_i$ = Jeb Bush, so measuring the Responsiveness between Trump and Bush over time period $h$ would also mean how much Donald Trump's influence on Jeb Bush to react on immigration is due to cues based on Mexican border.
\subsection{Multivariate Hawkes Process}
To model responsiveness, we borrow ideas from a general Point Process model called Hawkes Process. In particular, a multivariate Hawkes Process consists of $K$ point processes where the sequence of marked events $\{s_n,c_n\}_{n=1}^N$ have direct correspondence on the process in which the $n$-th event occurred. For each process $k$, it is a \textit{conditionally Poisson process} where there is an associated rate $\lambda_k(t | \{s_n:s_n<t\})$ based on the event history up to time $t$. 

We can further breakdown the rate function $\lambda_k(t)$ into two components: one based on the process $k$'s own history, also called the background rate function, in which we denote as $b_k(t | \{s_n:s_n<t \wedge c_n = k\})$. In this paper, we view $b$ as a constant throughout time, which is essentially the same as the background rate for Poisson process. 

The other is non-negative influence based on the event history of other point processes $k'$ within a time window of influence $\Delta t$, in which we denote as $$\rho_{k',k}(t | \{s_n: t-\Delta t< s_n < t\ \wedge (c_n=k \vee c_n=k')\}).$$ Figure 1 illustrates a sample Hawkes process. Whenever an event occurs at process $k$, we place a window of impulses on other processes $k'$ via $\rho_{k,k'}$ if $\rho_{k,k'}$ is nonzero. Superimposing the background rate $\lambda_k$ (denoted by flat horizontal line), we have a combined influence leading to event occurrence in any process $k$ over time.

In detail, we can further decompose $\rho_{k',k}(t)$ into $W_{k',k}\ g_{k',k}(t | \Delta t)$, where $W \in \mathbb{R}_{+}^{K \times K}$ is a non-negative weight matrix denoting influence from process $k'$ to $k$. $g$ is a convex combination of basis functions, each curtailed to be of length $\Delta t$ with nonzero values, so $g$ is a convolved signal in time with the basis functions.

The joint likelihood of the model is the product $$p(\{s_n,c_n\}_{n=1}^N | \lambda_k(t)) = \prod_{k=1}^K{p(\{s_n: c_n=k\} | b_k(t))}\prod_{n=1}^N\prod_{k=1}^K{p(\{s_n': c_n'=k\}|\rho_{c_n,k}(t-s_n))}.$$ 
\begin{figure}[t!]
\begin{center}
\includegraphics[scale=0.27]{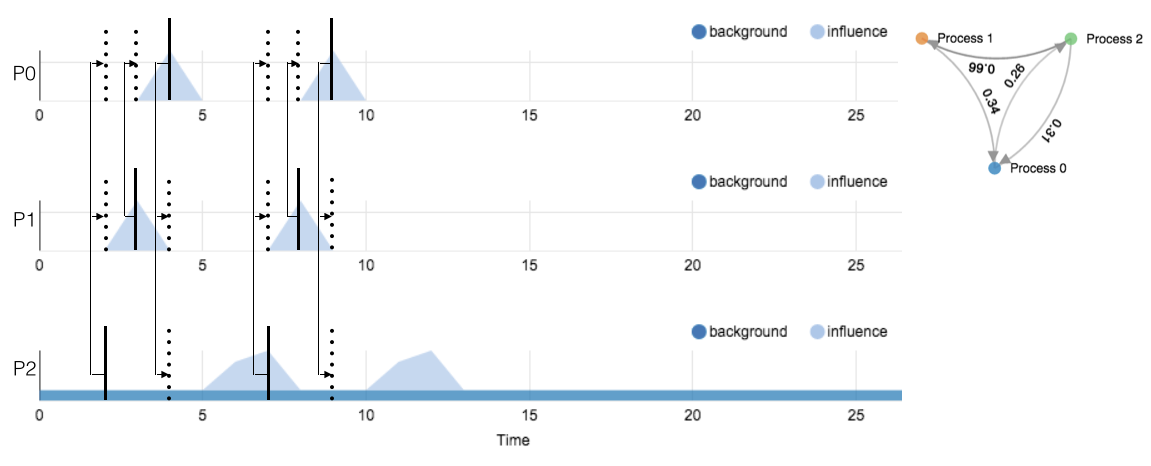}
\end{center}
\caption{A toy data showing influence rate function $\lambda(t)$. Solid line represents event occurrence; dotted line and light blue area represent influence due to $\rho(t)$. Top right shows the underlying influence network $W$.}
\end{figure}
For parameter learning, we use stochastic gradient descent on the log-posterior function, where we have placed a Gamma prior on the weight matrix. To use the outcome of the algorithm, we compute instantaneous responsiveness $R(s_i,t_i,t | C), t \in [0,T]$ to be 
$$R(s_i,t_i,t | C) = W_{s_i,t_i}\int_{\delta t=0}^{\Delta}{|\{s_n: c_n=t_i \wedge s_n=t-\delta t\}|\ g(t-\delta t) dt}$$
To represent responsiveness over time interval $h$, we simply integrate over instantaneous responsiveness rate and normalize by receiver $t_i$. 
\begin{figure}[t!]
\begin{center}
\includegraphics[scale=0.3]{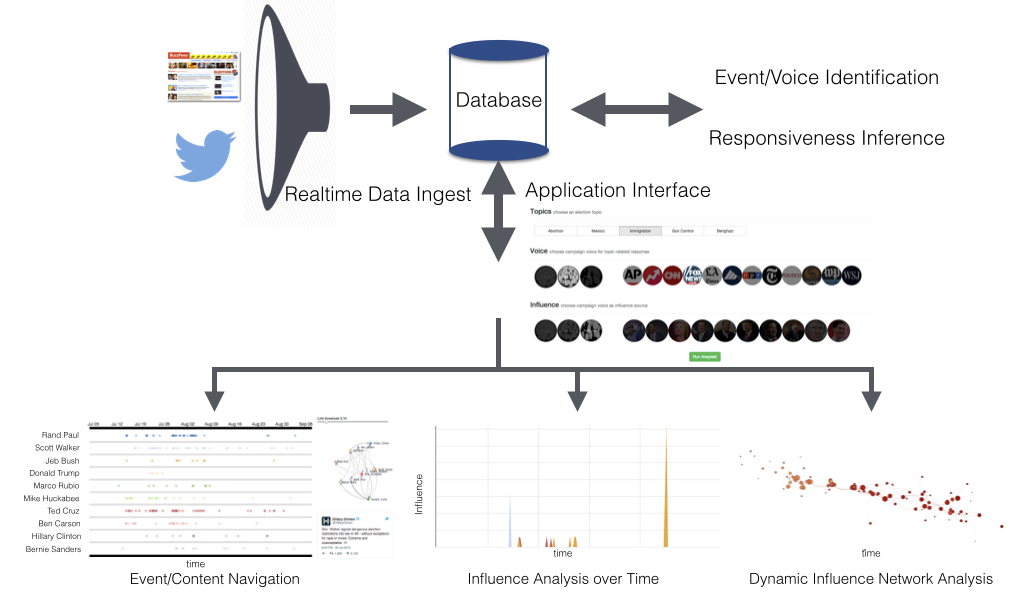}
\end{center}
\caption{System flowchart and snapshots of initial version interface.}
\end{figure}
\subsection{Beyond Recency: Modeling Responsiveness Attribution}
As mentioned earlier, Responsiveness can be a result of nuance signals. These signals may be terms used by a certain candidate (e.g. "Big Mexican Wall" remark from Donald Trump). As Hawkes Process mainly focuses on modeling the \textit{recency} effect, we need to augment the current model to have the ability to discerning effects due to such nuance signals or simply recency alone. In general, we can view these signals as two types of features. One type of feature associates with properties due to the occurring events, for example, mentioning of a certain topic in a tweet. Another type of feature is dyadic interactions between pairs of voices, for example, @ mentions between candidates in tweets. 

A marked event $\{s_n,\ c_n\}$ is associated with features of cardinality $F$, $f_n: (f_{n1},\ldots,f_{nF})$ and binary dyadic interactions $d_n: (d_{c_n1},\ldots,d_{c_nK})$. To model feature attribution, we use \textit{softmax} function as follows:
$$\mathcal F_n(i) = \frac{e^{\theta_{c_ni}^T f_i}}{\sum_j{e^{\theta_{c_nj}^T f_j}}}\ $$ 
The choice of softmax follows the notion of sparse additive property associated with it. This means that the feature contributing to most of the influence will be greatly stressed by the function. For dyadic interactions, we establish additional influence variable $\omega$, with corresponding weights $W'$:
$$ \omega_{k',k}(t | \{d_n: t-\Delta t< d_n < t\ \wedge (d_n=(k', k))\})$$
Based on the above notions, expressed in the form of intensity function, we can demonstrate the combined effects to be at time $t$, process $k$, and feature $i$:
$$\lambda_{\mathcal F}(t, k, i) = b_k + \sum_{k'=1}^K \sum_{s_n, c_n = k'} \rho_{k', k}(t-s_n) * \mathcal F_{s_n}(i) + \sum_{k'=1}^K \sum_{s_n, c_n=k'\wedge\ d_{n,k}>0} \omega_{k', k}(t-s_n)$$
The generalized model negative log-likelihood is specified:
$$\mathcal L = p(\{s_n,c_n\}_{n=1}^N | \lambda_k(t), \mathcal F_n) + \sum_{n=1}^N\sum_{i=1}^K\sum_{j=1}^K p(\{s_n: c_n=j\}|\omega_{j, i}(t-s_n)) + \mathcal R(\rho, \omega, \theta)$$

where we have included a penalty function for the parameters as well to avoid overfitting the noise. The associated data likelihood for the modified Hawkes process can be shown to be a convex function. Therefore, we can optimize the parameters with standard LBFGS solver. The Responsiveness attribution $R(s_i,t_i,h,\gamma | C)$ for feature $\gamma$ can be calculated as.
\begin{equation}
\begin{split}
R(s_i,t_i,h| \gamma, C) =& W_{s_i,t_i}\int_{\delta t=0}^{\Delta}{|\{s_n: c_n=t_i \wedge s_n=t-\delta t\wedge \mathcal F_n(\gamma) > 0 \}|\ \mathcal F_n(\gamma)\ g(t-\delta t) dt} \nonumber \\
&+ (\{s_n, c_n\} * \omega_{s_i, t_i})(t)|_0^\Delta\nonumber
\end{split}
\end{equation}

where the latter term is the convolution between point process data and dyadic interaction influence. Notice that currently the convolving time signal $g$ is a set of basis functions, which is fixed in advance before training. We can instead use a learned triggering kernel using a approximated bound for the log-likelihood and use Euler-Lagrange equation for kernel learning[14]. However, due to the concern of scaling in later research stages as well as simplicity, we omit the capability for model to learn custom shapes of $g$. For this paper, we opt for simple convex combination of basis function, where an exponential function yields similar results. 
\subsection{Solution System Framework}
As we build towards an end-to-end application system using inferred responsiveness, we want a system from data ingest to output visualization. Figure 2 shows the current flowchart and application snapshot of system design. The first part of the system consists of realtime data ingest from Twitter and 13 mainstream news sources. An NLP post-processing component is used to filter tweets and news articles that are election relevant. The next part contains voice and topic identification to filter contents and involving voices by user's interest. 

The main component of the application is the inference of responsiveness. Based on the instantaneous responsiveness $R(s_i,t_i,t | C)$, we can plot influence of other voices on a voice of interest over time. We can also examine the underlying weight matrix $W$ as an influence network. To check the exact cues that lead to influence, we enable the functionality to examine actual contents associated with the events.
\begin{figure}[t!]
\begin{center}
\includegraphics[scale=0.30]{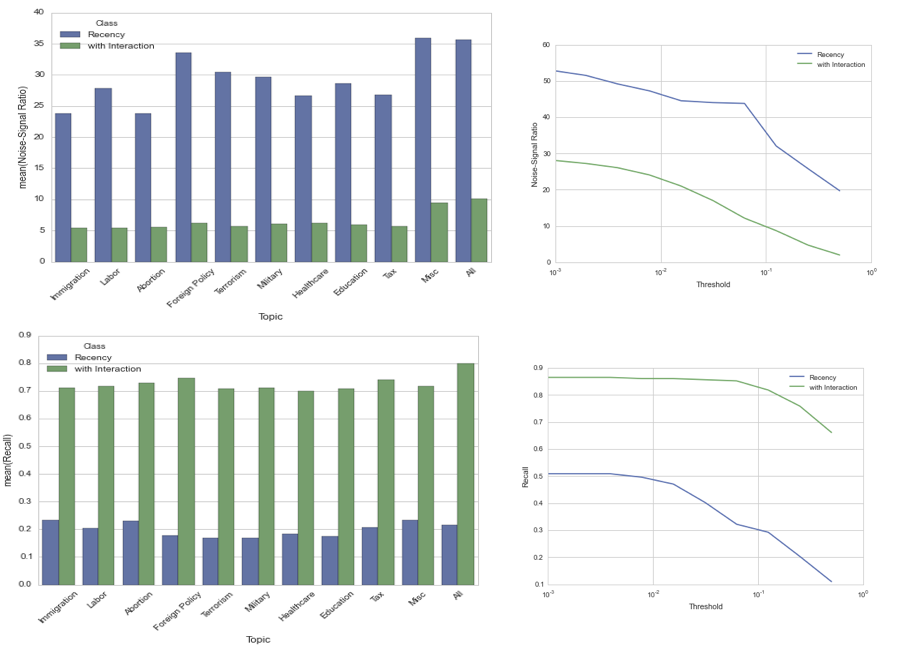}
\end{center}
\caption{Left: quality of inferred patterns over different topics; Right: over different selection thresholds.}
\end{figure}
\section{Related Work}
There are two lines of related work for measuring responsiveness in the online public sphere: analysis of influence and dialogue from political theory; computational modeling of multiple time series influence.

\subsection{Political Influence Analysis}
The first line of work is about political discourse analysis in the political theory. Political discourse analysis looks at how uses of rhetoric, lexicon, expression structure, interactions among the actors can shape political narrative[3]. It serves as a qualitative framework for analysis, and less of a quantitative methodology for measuring influence. 

There have been a line of works looking into the media ecosystem involving the politicians, mass media, and the general public [7]. In the context of U.S. presidential election, the researchers looked into how the politicians relayed messages through televised programs and ads, as well as press releases. They measure the public response through polling data. The works serve mostly as post-hoc statistical analytics. This paper tackles a different goal: we aim to model the online public sphere using social media, and develop algorithms to extract ripple effects \textit{for} post-hoc analysis.

\subsection{Multiple Process Dependence Modeling}
Modeling dependence between multiple time series relates to the concept of Granger Causality[5]. Granger Causality is a regression-based hypothesis test of causality between pairs of time series. However, in general, we would like to understand how each pair of voice influence jointly relates to one another as well as the nuances leading to different influence effects, which require capabilities beyond simple hypothesis testing.

There are many established probabilistic models for time series data, and correlations among multiple time series. Coupled-HMM[2] is a discrete hidden Markov state model where an observation on a particular time series may be due to latent states of all series in the previous timestamp. This type of models does naturally consider the notion of influence. Furthermore, stochastic point processes like Hawkes Process and event processes account for excitatory influence as well as influence accumulation (analogy made through survival analysis). In light of these assumption differences, the latter class of models seem to make more sense. 

We can also view each time series as a discrete sequence of events. Recently, variants of Hawkes process have been proposed in the machine learning community to model latent dependencies among a set of points processes[1][8][13]. In particular, several works look into the area of how signals may be the cause to influence, rather than recency[10][11]. What emerge from these works is the phenomenon of competing or cooperating time series, which makes good analogy to responsiveness modeling. The Bayesian nature of the models and their flexibility for generative hypothesis make this family a good option for first-pass design.
\section{Initial Applications}
\subsection{Data Collection}
Based on the initially implemented version of application, we are taking the ingested data from the data pipeline for preliminary analysis. For news articles, we have a realtime article ingest of 13 mainstream news organizations(AP, Buzzfeed, CNN, Fox News, LA Times, McClatchy, New York Times, NPR, Politico, ProPublica, Reuters, Wall Street Journal, Washington Post). For tweets, we include all tweets posted by news organizations, presidential candidates, and any tweet with text matching a predefined query set. We use a particular set of topics: [abortion], [Mexican border], [immigration], [gun control], [Syria], and [terrorism]. 

In addition, to test the ability of causal pattern extraction, we constructed a set of events between pairs of candidates where we are certain of causality. The decision criterion is that a candidate posts a direct response to another candidate on an election topic because that candidate mentioned her in a previous tweet. Based on a set of 15 candidates placed on top of the poll on Sep. 1, 2015, we extracted 878 distinct event sequences of which 304 the causality is certain.
\subsection{Validating Inferred Influence}
Here we run two versions of the algorithm on the candidate interaction data, one only based on recency, and another based on directed interactions. We measure the inferred rates from the learned models. We run the models on each topic and all topics jointly. To measure the quality of inferred influence, we compare with ground truth based on two measures: \textit{recall} where we want to cover important patterns; \textit{noise-signal ratio}, defined as |significant patterns|/|correctly inferred patterns| as opposed to precision because the tool aims to be explorative. We show the results in Figure 3. The model that considers interactions shows clearly higher recall and low noise ratio, with reasonable threshold. We believe considering detailed features as part of future work can further improve the system quality. 
\subsection{Influence Analysis}
For preliminary analysis, we test on three potential applications: interest alignment analysis, situational analysis, and topic shift analysis.
\subsubsection{Candidate-Candidate Influence and Alignment of Interest}
\begin{figure}[t!]
\begin{center}
\includegraphics[scale=0.30]{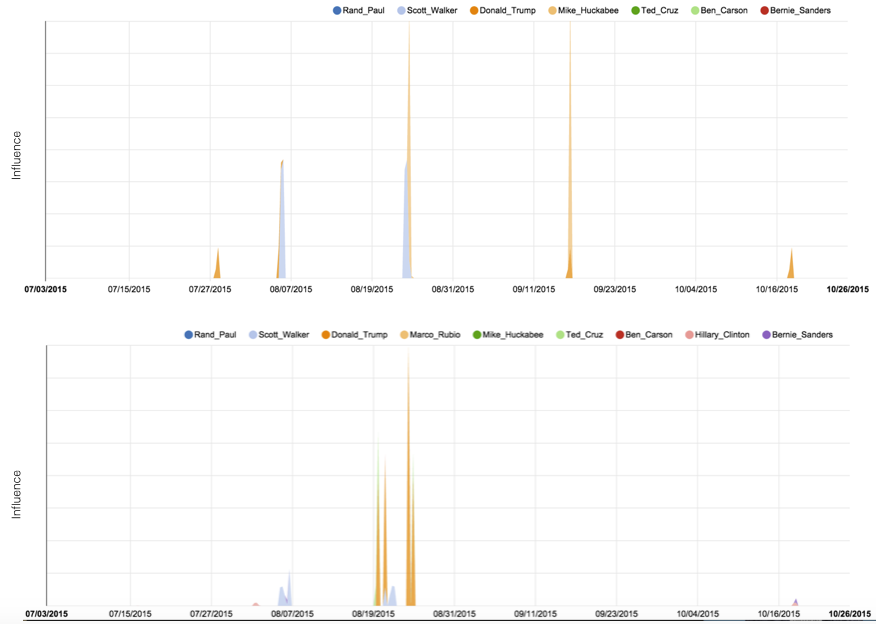}
\end{center}
\caption{Inferred responsiveness showing the influences of other presidential candidates on Jeb Bush, making Jeb Bush tweeting about \textit{Mexico}(top), \textit{immigration}(bottom) at given time.}
\end{figure}
If we let the voices $s_i,\ t_i$ each represented by a presidential candidate, we can use the candidate-to-candidate responsiveness to understand topic interest alignment between the candidates. Figure 4 shows inferred responsiveness over time for presidential candidates influencing Jeb Bush to tweet about \textit{Mexican border} and \textit{immigration}. When we navigate through the time (x-axis), we see spikes on the y-axis. These spikes represent strong influences from certain candidates. If we are interested in how Donald Trump influences Jeb Bush's narrative, we see that this is only the case for immigration, whereas Trump has very little influence on Jeb Bush over Mexico debate. This reveals an interesting phenomenon of the immigration debate among the Republican candidates: because Donald Trump was making inflammatory comments about Mexican immigrants, other candidates were consciously avoiding Trump's narrative on Mexico and they instead focused more on the general immigration debate.
\subsubsection{Media's Preference on Candidates}
It is also interesting to look at the responsiveness of news media influenced by presidential candidates' comments on Twitter. Figure 5 shows the candidate influence on Fox News to post articles on \textit{Mexican border} and \textit{immigration} respectively. We can observe the contrast how Donald Trump strongly influenced Fox to write about both Mexican border and immigration. Also, we can see how Fox news lowered its interest on the both topics after September began.
\begin{figure}[t!]
\begin{center}
\includegraphics[scale=0.25]{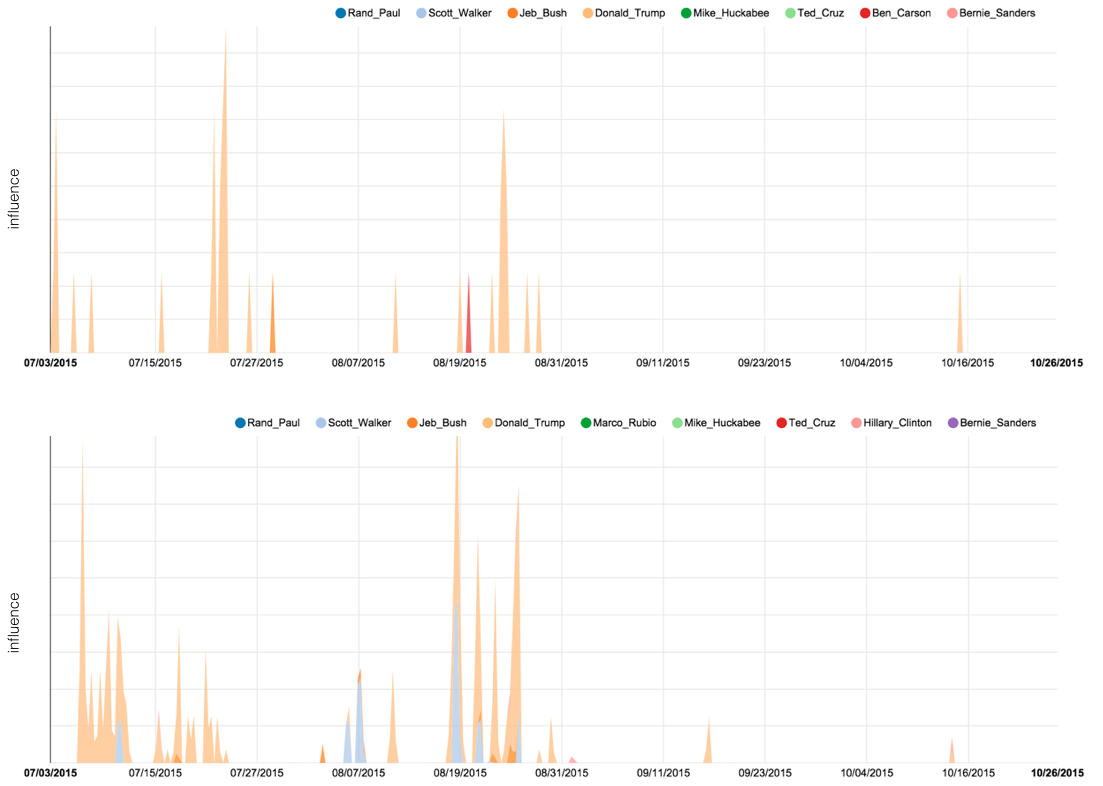}
\end{center}
\caption{Inferred responsiveness showing the influences of presidential candidates on Fox News writing about \textit{Mexico}(top), \textit{immigration}(bottom) at given time.}
\end{figure}
\subsubsection{Situational Analysis}
Lastly, the responsiveness can be used for situational analysis to compare the differences of aligned interest before and after a certain news event. Take Figure 6 (left) as an example, the chord diagram shows the candidate influence on Hillary Clinton tweeting about abortion before and after the first Republican debate. Before the debate, we see a big proportion of influence from Rand Paul and Scott Walker, who were both very active on the topic. However, after the debate, there was a big drop of influence from the two candidates. If we referred to the Republican poll[9], we knew the two candidates's support dropped substantially post debate. It seemed natural for Hillary Clinton to shift focus towards more competitive Republican candidates on the issue. 

Another example of situational analysis is the analysis on shift in topical influence over time. Figure 6 (right) displays how the topical influences that led to the differences in news media's coverage on \textit{immigration} varied before and after the beginning of September. Interestingly, the media coverage on Mexican border dropped significantly after the Republican candidates visited the border in late August. Therefore, there was less influence from the topic Mexican border on immigration starting from September. 
\begin{figure}[h]
\begin{center}
\includegraphics[scale=0.25]{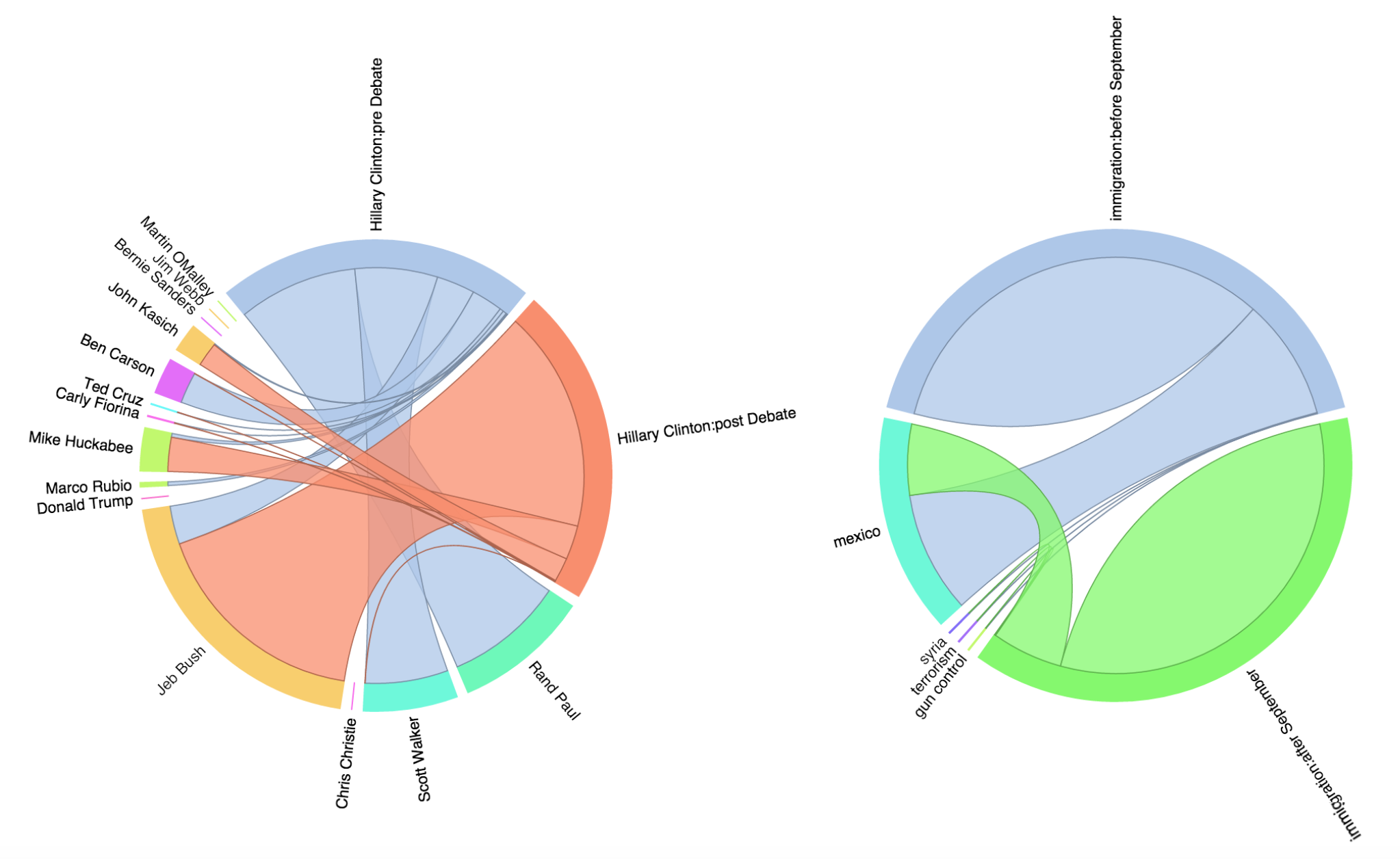}
\end{center}
\caption{Illustration of (left) candidate influence on Hillary Clinton tweeting about abortion before and after the first Republican debate; (right) topical influence on news media writing about immigration before the after the beginning of September.}
\end{figure}
\section{Discussions and Future Work}
In this paper, we have proposed to use multivariate Hawkes Process to model influence dependence between different voices, or point processes, over time. We have implemented an initial version of the end-to-end system that takes the ingested data for influence analysis in the online public sphere. We demonstrate a few use cases where our responsiveness can deliver interesting analysis, where the outcomes may be used as part of journalistic reports discussing the opinion influence dynamics in the public social media ecosystem. 

Although the preliminary analysis suggests some interesting potential application using the responsiveness measure, this work is still very preliminary in general. There are several important hanging challenges that need to be resolved in order to make the output of the responsiveness measure more useful:
\begin{itemize}
\item The current model is not very good at handling mixed types of voices. For example, the sequence on general public tweeting about a topic will have much higher magnitude than presidential candidates. How can we unify the effect of popularity in each voice, beyond simple normalization, so the inferred responsiveness is not easily dominated by the difference in scale (by placing assumption on virality of a topic, as described in [12])?
\item Generating a clean set of ripple effect patterns comes down to an old problem in information retrieval: how can we balance between precision and recall? How do sparse models affect pattern quality?
\item On individual level, event occurrence may be attributed to duration modeling, so responsiveness may be due to either direct influence or cumulating susceptibility. How can we model these two disparate effects jointly? The question warrants further investigation.
\item Dive deeper into features, topics, and stylistic differences. Can we consider text as part of generative process? Understanding how expressions coevolve with influence may reveal additional interesting patterns.
\end{itemize}

\subsubsection*{Acknowledgements}
The authors would like to thank Sophie Chou, Neo Mohsenvand, Prashanth Vijayaraghavan, and Soroush Vosoughi for suggestions on research, data collection, and providing article data. The authors also thank Ryan Adams, Allen Gorin, and Scott Linderman for providing resources and suggestions on algorithm design.
\subsubsection*{References}

\small{
[1] Blundell, Charles, Heller, Katherine, and Beck, Jeffery. (2012) Modelling reciprocating relationships with Hawkes processes. \textit{Advances in Neural Information Processing Systems}, 2012.

[2] Brand, M., Oliver, N., Pentland, Alex (1997) Coupled hidden Markov models for complex action recognition, Proceedings of the IEEE CS Conference on Computer Vision and Pattern Recognition, 1997, pp. 994?999.

[3] Chilton, Paul. (2004) Analyzing Political Discourse Theory and Practice. Routledge, Taylor \& Francis Group, London and New York.

[4] Dennett, Daniel \& Roy, Deb. (2015) How Digital Transparency Became a Force of Nature. In Scientific American Volume 312, Issue 3. 

[5] Granger, C. W. J. (1969). Investigating Causal Relations by Econometric Models and Cross-spectral Methods. Econometrica 37 (3): 424?438. 

[6] Habermas, Jürgen (1962), The Structural Transformation of the Public Sphere: An Inquiry into a Category of Bourgeois Society, Thomas Burger, Cambridge Massachusetts: The MIT Press.

[7] Just, Marion, Crigler, Ann, Alger, Dean, Cook, Timothy (1996). Crosstalk: Citizens, Candidates, and the Media in a Presidential Campaign. University of Chicago Press.

[8] Linderman, Scott \& Adams, Ryan. (2014) Discovering latent network structure in point process data. In \textit{Proceedings of The 31st International Conference on Machine Learning}, pages 1413-1421, 2014.

[9] RealClearPolitics - Election 2016 - 2016 Republican Presidential Nomination. $www.realclearpolitics.com/epolls/2016/president/us/2016\_republican\_presidential\_nomination-$ $3823.html$, accessed November 2, 2015.

[10] Valera, Isabel, Gomez-Rodriguez, Manuel (2015) Modeling Adoption and Usage of Competing Products. IEEE International Conference on Data Mining (ICDM 2015), Atlantic City, 2015.

[11] Zarezade, Ali et al. (2016) Correlated Cascades: Compete or Cooperate. WSDM, 2016.

[12] Zhao, Q., Erdogdu, M., He, H., Rajaraman, A., and Leskovec, J. (2015) SEISMIC: A Self-Exciting Point Process Model for Predicting Tweet Popularity. ACM SIGKDD International Conference on Knowledge Discovery and Data Mining (KDD), 2015. 

[13] Zhou, Ke, Zha, Hongyuan, and Song, Le. (2013) Learning social infectivity in sparse low-rank networks using multi-dimensional Hawkes processes. In \textit{Proceedings of the International Conference on Artificial Intelligence and Statistics}, volume 16, 2013.

[14] Zhou, Ke, Zha, Hongyuan, and Song, Le. (2013) Learning Triggering Kernels for Multi-dimensional Hawkes Processes. In \textit{Proceedings of the 30th International Conference on Machine Learning}, Atlanta, Georgia, 2013.
%
%
}

\end{document}